\documentclass{article}
\usepackage{ltwol2e}

 \usepackage{epsfig}

\def\Journal#1#2#3#4{{#1} {\bf #2}, #3 (#4)}

\def\NIMA{{\em Nucl. Instrum. Methods} A}

\def\be{\begin{equation}}
\def\ee{\end{equation}}
\def\bea{\begin{eqnarray}}
\def\eea{\end{eqnarray}}

\begin{document}

\pagestyle{empty}

\begin{titlepage}

\begin{flushright}
{\normalsize
SLAC--PUB--8017\\
November 1998\\}
\end{flushright}

\vspace{2.5 cm}

\begin{center}
      {\large\bf Measurements of $A_{LR}$ and $A_{lepton}$ from SLD}
\end{center}

\vspace{2.0 cm}

\begin{center}
        K. G. Baird \\
	Department of Physics\\
        University of Massachusetts, Amherst, MA 01003, USA \\ ~~\\  ~~\\
        Representing the SLD Collaboration \\
        Stanford Linear Accelerator Center \\
        Stanford University, Stanford, CA 94309, USA \\
\end{center}

\begin{center}

\vspace{15mm}
 {\bf Abstract}
\vspace{5mm}

\end{center}

{\normalsize
\noindent
This paper presents measurements of the leptonic asymmetries in $Z^0$
decays measured with the SLD detector. Using a data sample of
approximately 500,000 $Z^0$ bosons, we report preliminary values for
$A_e$, $A_{\mu}$, and $A_{\tau}$ using both hadronic and leptonic
decays. When combining all results, we report a preliminary value
for the effective weak mixing angle 
$\sin^2 \theta_W^{\sl eff}=0.23110 \pm 0.00029$.
}

\begin{center}

\vspace{45mm}
{\normalsize\sl
     Presented at the XXIX$^{th}$ International Conference on High Energy
     Physics,
     23-29 July 1998, Vancouver, Canada.}

\vspace{15mm}
{\footnotesize Work supported in part by Department of Energy Contracts
                DE--AC03--76SF00515(SLAC) and 
		DE-FG02-92ER40715(Massachusetts).}
\end{center}

\pagestyle{plain}

\end{titlepage}

\title{Measurements of $A_{LR}$ and $A_{lepton}$ from SLD}

\author{K. G. Baird}

\address{Department of Physics, University of Massachusetts, Amherst,
Mass. 01003, USA\\ E-mail: baird@slac.stanford.edu}   


\twocolumn[\maketitle\abstracts{ 
This paper presents measurements of the leptonic asymmetries in $Z^0$
decays measured with the SLD detector. Using a data sample of
approximately 500,000 $Z^0$ bosons, we report preliminary values for
$A_e$, $A_{\mu}$, and $A_{\tau}$ using both hadronic and leptonic
decays. When combining all results, we report a preliminary value
for the effective weak mixing angle 
$\sin^2 \theta_W^{\sl eff}=0.23110 \pm 0.00029$.}]

\section{Introduction}

The first-of-its-kind SLC linear electron positron collider~\cite{slc} has
proven to be a powerful facility for tests of the Standard Model
via the measurement of Electroweak couplings at the $Z^0$ pole.
Its highly longitudinally-polarized electron beam
($P_e \sim 75\%$) and small luminous region of
($1.5 \times 0.8 \times 700$) $\mu$m in (x,y,z) are particularly
advantageous for the measurement of electroweak quantities.
The SLD detector, described in more detail elsewhere~\cite{sld}
is tailored to take maximum advantage of these
attributes of the SLC.

\subsection{Asymmetries}\label{subsec:asym}

 At Born level, for an electron beam polarization $P_e$,
the differential cross section for the process \\
$e^+ e^- \rightarrow Z^0 \rightarrow f {\overline f}$ is given by
\begin{eqnarray}
 \sigma^f (z) & \propto & [(v_e^2 + a_e^2 - 2a_e v_e P_e)
(v_f^2 + a_f^2)](1+ z^2) \nonumber \\
              & + &[2a_e v_e - (v_e^2 + a_e^2)P_e]4 v_f a_f z
\label{eq:murnf}
\end{eqnarray}
where $z = \cos \theta$ is the angle of the final state fermion with respect
to the beam axis, and $\sigma^f(z) = d\sigma_f/dz$.
Here, it is assumed that $P_e = +1$ for a right-handed
(positive helicity) electron beam.
Thus, the vector (v) and axial-vector (a) couplings are
the free parameters which specify the $Z-f$ coupling. In the Standard Model,
\begin{equation} 
v = I_3^{Weak} - 2Q\sin^2\theta_W \label{eqn:v}
\end{equation}
\begin{equation} 
a = I_3^{Weak}.   \label{eqn:a}        
\end{equation}
With the use of a {\it polarized} electron beam, these coupling parameters
can be extracted independently of the initial state couplings
$v_e$ and $a_e$, via the two following observables. The first
of these, the ratio of partial widths
\begin{equation} 
R_q \equiv {\int_{-1}^{1} \sigma_q(z)dz \over
   \sum\limits_{q'} \int_{-1}^{1} \sigma_{q'}(z)dz}   =
  {v_q^2 + a_q^2 \over \sum\limits_{q'}(v_{q'}^2 + a_{q'}^2)} =
   {\Gamma_q \over \Gamma_{had}},   \label{eqn:Rq} 
\end{equation}
does not require the use of polarized beams; 
the restriction of the denominator
to quark species only (no leptons) leads to the 
cancellation of QCD radiative effects.
The second of these, the polarized forward-backward asymmetry,
does require the use of polarized beams:
\begin{eqnarray}
{\tilde A}_{FB}^f(z) & = &
{[\sigma^f_L(z) - \sigma_L^f(-z)] - [\sigma_R^f(z) - \sigma_R^f(-z)] \over
  \sigma^f_L(z) + \sigma_L^f(-z)  +  \sigma_R^f(z) + \sigma_R^f(-z)  }
\nonumber \\
& = & |P_e|A_f {2z \over 1 + z^2},  \label{eqn:ALRFB} 
\end{eqnarray}
where the subscript L (R) refers to left-handed (right-handed) electron
beam, and
\begin{equation} 
A_f \equiv {2 v_f a_f \over  v_f^2 + a_f^2}   \label{eqn:Af}   
\end{equation} 
is the quantitative extent of parity violation in the $Z^0f\bar{f}$ coupling.
These two measurements specify the $Z^0f\bar{f}$ coupling in generality.

Finally, one can define the left-right asymmetry
\begin{eqnarray}
A_{LR} & = &{1 \over P_e} {\sum\limits_{f \ne e} \int_{-1}^{1} \sigma_L^f(z) dz -
             \sum\limits_{f \ne e} \int_{-1}^{1} \sigma_R^f(z) dz \over
             \sum\limits_{f \ne e} \int_{-1}^{1} \sigma_L^f(z) dz +
             \sum\limits_{f \ne e} \int_{-1}^{1} \sigma_R^f(z) dz }
             \nonumber \\
 & = &
 {1 \over P_e} [P_e {2 v_e a_e \over v_e^2 + a_e^2}] = A_e  \label{eqn:alr}
\end{eqnarray}
where the sum is restricted to all final states except
$e^+e^-$ in order to avoid having to unravel t-channel effects.
This is a particularly potent way to measure $\sin^2 \theta_W$ provided
$P_e$ can be measured precisely:
\begin{equation} 
A_{LR} = {2(1 - 4\sin^2\theta_W) \over 1 + (1 - 4\sin^2 \theta_W)^2}
\label{eqn:alreff} 
\end{equation} 
making $A_{LR}$ very sensitive to the weak mixing angle:
\begin{equation} 
{d A_{LR} \over d \sin^2 \theta_W} \simeq -7.8. \label{eqn:alrsen} 
\end{equation}

It should be pointed out that these relations have been derived for
the case of the Born-level interaction. However, since the
$Z^0$-pole measurements now provide the most accurate
constraints on Standard Model consistency, the convention that
has arisen is to incorporate higher order effects by making
Eqn.~\ref{eqn:alreff} the {\it definition} of the weak mixing angle.
This is denoted by the notation $\sin^2\theta_W^{\sl eff}$; higher
order effects must then be explicitly accounted for when
comparing this value with that of non-$Z^0$-pole measurements.

\begin{figure*}
\psfig{figure=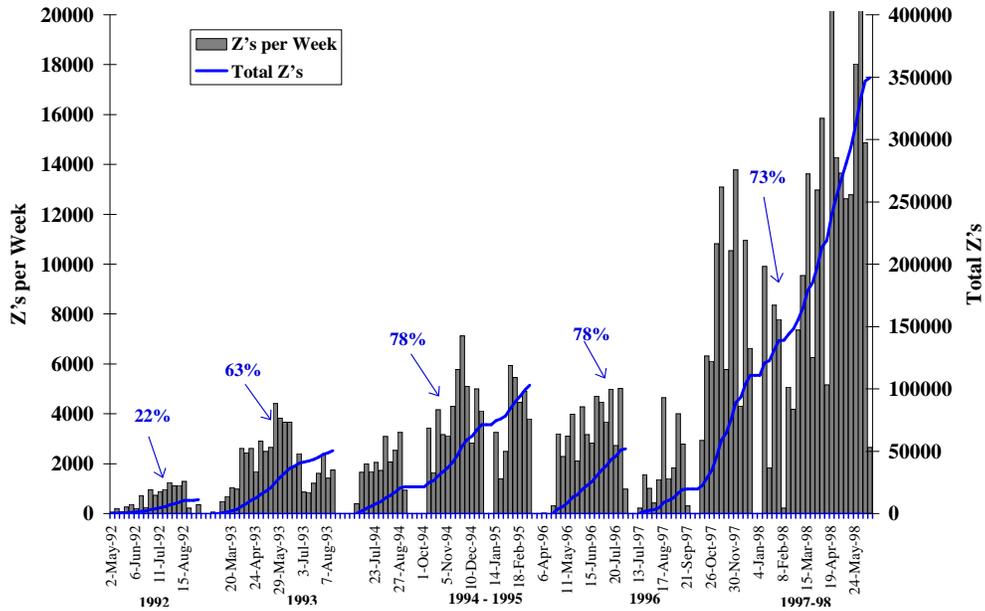,width=15cm}
\caption{Luminosity history of the SLD experiment, along with average
beam polarizations for each run.}
\label{fig:lumi}
\end{figure*}

In this paper, we discuss the most recent measurements of $A_{LR}$ and
$A_{e-\mu-\tau}$ (or $A_{lepton}$) performed by the SLD Collaboration.
Both analyses utilize the approximately 350K $Z^0$ decays which were 
obtained in the 1997-98 physics run, and
combine these new results with those from the earlier dataset of 
approximately 200K $Z^0$ decays (see Fig. \ref{fig:lumi}).
After first discussing polarimetry,
which is crucial for both analyses, we discuss the measurement
of $A_{LR}$ and the corrections need to express it in terms of the
$Z^0$-pole asymmetry shown in Eqn. \ref{eqn:alreff}. We then present
measurements of $A_{lepton}$ using final-state leptons. We conclude by
combining our results for $\sin^2\theta_W^{\sl eff}$, and comparing our
results with leptonic and hadronic measurements from LEP.

\vspace*{-1.8pt}   

\section{Polarization Measurements}
\begin{figure*}
\psfig{figure=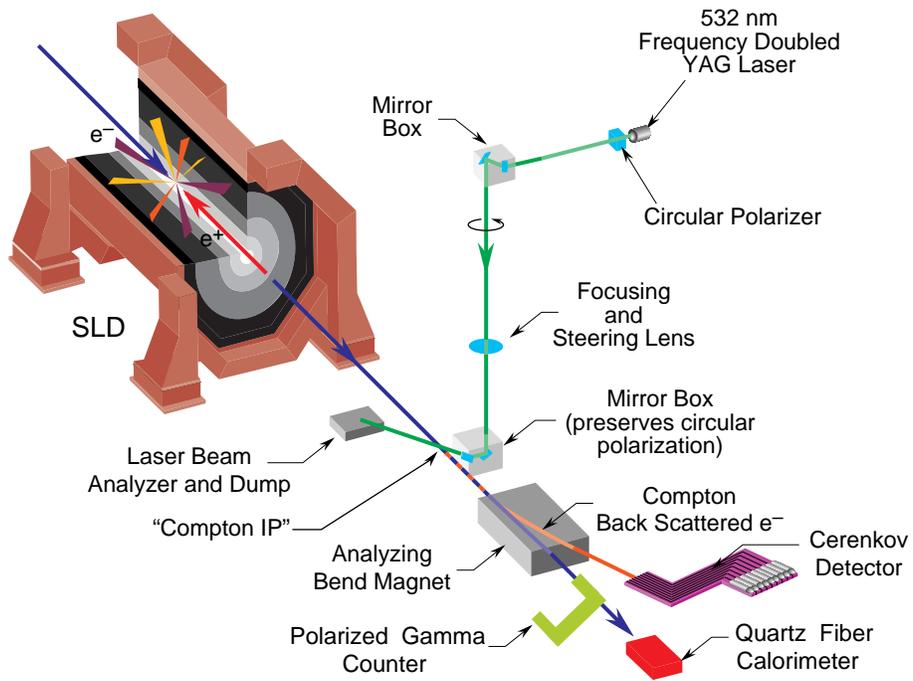,height=9cm}
\caption{A view of the SLD Polarimeters used to determine
electron-beam polarization. The two new devices, the Polarized Gamma
Counter and the Quartz Fiber Calorimeter, examine the Compton
backscattered photons.}
\label{fig:polarimeters}
\end{figure*}

Precision polarimetry of the SLC electron beam is accomplished
with the Compton Polarimeter~\cite{compton}, which employs Compton scattering
between the high-energy electron beam and a polarized
Nd:YAG laser beam ($\lambda = 532$nm) to probe the
electron beam polarization. The Compton scattered electrons, 
which lose energy in the scattering process
but emerge essentially undeflected, are analyzed by the
first beam-line dipole downstream of the SLD interaction point (the 
``Analyzing Bend Magnet'' shown in Fig. \ref{fig:polarimeters}),
beyond which they exit the beam-line vacuum through a thin window,
and enter a threshold Cerenkov detector segmented transverse to the
beam-line.

The laser beam polarization, typically 99.9\%, is continuously
monitored. The average polarization of each bunch is measured, but the
relevant quantity for analysis is the luminosity-weighted polarization
at the IP. This leads to a very small correction ($< 0.1\%$), whose
uncertainty is shown in in Table  \ref{table:polarization}. 
Also shown is this table is the luminosity-weighted average
polarization for each run, as well as the relative systematic 
uncertainties in the Compton
polarization measurement. In 1996 a relative polarization 
scale uncertainty of 
$\pm 0.64$\% was obtained; the preliminary 1997/98 values of $\pm
1.03$\% will improve substantially once the analysis is complete.

In addition to the Compton Polarimeter,
two additional detectors are in place to measure the electron beam
polarization by examining the Compton backscattered photons. These two
devices~\cite{pgc-qfc}, the Polarized Gamma Counter (PGC) and 
the Quartz Fiber Calorimeter (QFC), both require dedicated
electron-only conditions due to beamstrahlung backgrounds. Both
devices provide sub-1\% cross-checks of the Compton polarimeter.

\begin{table*}
\caption{ The
year-by-year luminosity-weighted average polarization for each run
(with statistical error),
along with the systematic uncertainty on the electron beam polarization
scale, as measured by the Compton Polarimeter. The 1996-98 results
are preliminary, and  the systematic uncertainties are expected to be
reduced after the full analysis is completed.}
\label{table:polarization}
\vskip6pt
\begin{center}
\begin{tabular}{|l|c|c|c|c|c|c|}
\hline

                           & 1992   & 1993 & 1994/5 & 1996* &  1997* & 1998*\\
{\bf Ave. Polarization}&{\bf $0.224\pm0.006$ }& {\bf
$0.626\pm0.012$}&{\bf $0.772\pm0.005$} 
                           &{\bf $0.765\pm0.005$} & {\bf $0.733\pm0.008$} &{\bf $0.731\pm0.008$} \\
\hline
\hline
Sys. Uncertainties&        &           &            &       &        &      \\
\hline
Laser Polarization   &   2.0  &  1.0 &  0.20  &  0.20 &  0.20  &  0.10\\
Detector Linearity   &   1.5  &  1.0 &  0.50  &  0.50 &  0.50  &  0.50\\
Detector Calibration &   0.4  &  0.5 &  0.29  &  0.30 &  0.30  &  0.30\\
Electronic Noise     &   0.4  &  0.2 &  0.20  &  0.20 &  0.20  &  0.20\\
Interchannel Consist. &   0.9 &  0.5 & --- & ---  &  0.80  &  0.80\\[3pt]
\hline
Sum of Pol. Unc.  &   2.7 &  1.6 & 0.64 & 0.64 & 1.03  & 1.01 \\[3pt]
\hline
Compton/SLD IP           &   --- &  1.1 & 0.17 &  0.18 & 0.07 &  0.08\\ [3pt]
\hline
\hline
{\bf Total $P_e$ Unc.} & {\bf 2.7\%}
& {\bf 1.9\%} & {\bf 0.67\%} & {\bf 0.67\%}   & {\bf 1.03\%}& {\bf 1.01\%}        \\
\hline
\end{tabular}
\end{center}
\end{table*}

\section{The Measurement of $A_{LR}$}
In practice, $A_{LR}$ is measured with hadronic final states
only, as the selection efficiency for $\tau^+\tau^-$ and 
$\mu^+\mu^-$ pairs is very low (the leptonic final states are 
considered separately). For the last three measurements (1996, 1997 and 1998),
a 99.9\% pure hadronic sample was selected by requiring that the 
absolute value of the energy imbalance (ratio
of vector to scalar energy sum in the calorimeter) be
less than 0.6, that there be at least 22 GeV of visible
calorimetric energy, and that at least 3 charged tracks be reconstructed
in the central tracker; this selection was 92\% efficient. After
counting the number of hadronic decays for left- and right-handed
electron beams and forming a Left-Right Asymmetry, a small
experimental correction for
backgrounds (and negligible corrections for false asymmetries) was
applied; for the 1998 dataset this correction was 0.06\%. This
then yielded a value for the measured Left-Right Asymmetry (1998) of
\begin{eqnarray}
A_{LR}^{meas}& = & {1 \over P_e}
{N_L - N_R \over N_L + N_R} \\
& = & 0.1450 \pm 0.0030 \pm 0.0015 
\label{eqn:alrmeas}
\end{eqnarray}
where the systematic uncertainty is dominated by the uncertainty in
the polarization scale. The translation of this result to the  
$Z^0$-pole asymmetry $A_{LR}^0$ was a $1.8 \pm 0.4$\% effect, where
the uncertainty arises from the precision of the center-of-mass energy
determination. This small error due to beam energy uncertainty is
slightly larger that seen previously (it has been quoted as $\pm
0.3$\%), and reflects the results of a scan of the Z peak used to
calibrate the energy spectrometers to LEP data, which was performed
for the first time during the 1998 run. This correction 
yields the 1998 preliminary result of
\begin{eqnarray}
  A_{LR}^0 & = &0.1487 \pm 0.0031 \pm 0.0017 \label{eqn:1998alr0}\\
  \sin^2\theta_W^{\sl eff}& =& 0.23130 \pm 0.00039 \pm 0.00022.
\label{eqn:1998sin2tw}
\end{eqnarray}
As an additional crosscheck, in 1998 a dedicated experiment was
performed in order to directly test the expectation that any accidental
polarization of the positron beam is negligible. The Moller
Polarimeter~\cite{band_prepost} in SLAC End Station A was used to
analyse the bhabha
process for the $e^+$ beam; the (preliminary) results of this
measurement indicated
that the $e^+$ polarization was consistent with zero: $P_{e^+} = -0.02
\pm 0.07\%$.

\begin{figure}
\center
\psfig{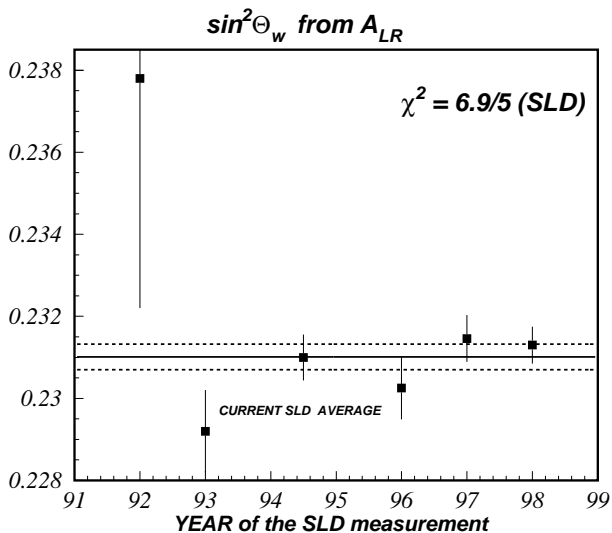}
\caption{The six measurements of $A_{LR}^0$ performed by the SLD
 experiment. Here, the results have been translated into
 $\sin^2\theta_W^{\sl eff}$ using Eqn. \ref{eqn:alreff} in the text. }
\label{fig:9298sin2tw}
\end{figure}

The six measurements $A_{LR}^0$ performed by SLD are shown in
Table \ref{tab:9298sin2tw}, along with their translations to 
$\sin^2\theta_W$. In Figure \ref{fig:9298sin2tw} the six
$\sin^2\theta_W$ results are plotted as a function of measurement
year; clearly the measurement fluctuations are statistical in nature.
The (preliminary) averaged results for 1992-98 are
\begin{eqnarray}
A_{LR}^0 = 0.1510 \pm 0.0025 
\label{eqn:9298alr0}\\
\sin^2\theta_W^{\sl eff} = 0.23101 \pm 0.00031
\label{eqn:9298sin2tw}
\end{eqnarray}

\begin{table*} 
\caption{\label{tab:9298sin2tw}
The six measurements of $A_{LR}^0$ and $\sin^2\theta_W^{\sl eff}$ 
performed by the SLD  experiment.}
\vskip6pt
\begin{center}
\begin{tabular}{|l|lllll|lll|}
\hline 
{Run} &\mbox{ } &{$A_{LR}^{0}$} &\mbox{ } & $\hspace{0.0in}\delta A_{LR}^{0}$& \mbox{ }&
{$\sin^{2}\theta_{w}^{\mbox{\tiny eff}}$} & \mbox{ } &$\hspace{0.0in}\delta \sin^{2}\theta_{w}^{\mbox{\tiny eff}} $\\
\hline
{1992} & &{0.100} & &$\pm 0.044 \pm 0.004 $ & &{0.2378} & &$\pm 0.0056 \pm 0.0005$ \\
{1993} & &{0.1656}& &$\pm 0.0071\pm 0.0028 $ & &{0.2292} & &$\pm 0.0009 \pm 0.0004$ \\
{1994-95} & &{0.1512}& &$\pm 0.0042 \pm 0.0011$& &{0.23100}& &$\pm 0.00054 \pm 0.00014$ \\
{1996$^*$} (preliminary)& &{0.1570}& &$\pm 0.0057\pm 0.0011$& &{0.23025}& &$\pm 0.00073\pm 0.00014$ \\ 
{1997$^*$} (preliminary)& &{0.1475}& &$\pm 0.0042\pm 0.0016$& &{0.23146}& &$\pm 0.00054\pm 0.00020$ \\ 
{1998$^*$} (preliminary)& &{0.1487}& &$\pm 0.0031\pm 0.0017$& &0.23130& &$\pm 0.00039\pm 0.00022$ \\
\hline \hline
Combined & &{0.1510} & & {$\pm$ 0.0025}& &{0.23101}& &{$\pm$ 0.00031} \\ \hline
\end{tabular}
\end{center}
\end{table*}

\section{The Measurements of $A_e$, $A_\mu$,$A_\tau$ using 
final-state leptons}

Parity violation in the $Z^0$-lepton couplings is measured
by fits to the differential cross section (Eqn. \ref{eq:murnf}) separately
for the three leptonic final states, including the effects of
t-channel exchange for the $Z^0 \rightarrow e^+ e^-$ final
state. For the 1996-98 data sample, leptonic final states in the range
$|\cos\theta| < 0.8$ are selected by requiring that events have
between two and eight charged tracks in the CDC, at
least one track with $p \ge 1$ GeV/c. One of the two event hemispheres
was required to have a net charge of -1, while the other one had to
have a net charge of +1. 
Events are identified as bhabha candidates if
the total calorimetric energy associated with the two
most energetic tracks is greater than 45 GeV. On the
other hand, if there is less than 10 GeV of associated energy
for each track, and there is a two-track combination with
an invariant mass of greater than 70 GeV/c$^2$, the event
is classified as a muon candidate. The selection criteria for tau
candidates is somewhat more complex. First, the $\tau^+\tau^-$ pair
invariant mass is required to be less than 70 GeV/c$^2$, 
with a separation angle between the two tau momentum vectors 
of at least 160$^{\circ}$. At least one track in the event must have a
momentum greater than 3 GeV/c. Each $\tau$ hemisphere must have an
invariant mass less that 1.8 GeV/$c^2$ (to suppress $Z^0$ hadronic
decays), and the associated energy in the LAC from each track must be
less than 27.5(20.0)~GeV for tracks with polar angles less than
(greater than) $|\cos\theta| = 0.7$. 
The resulting efficiencies and purities are shown in Table 
\ref{table:leptonstats}.

\begin{table*}
\label{table:leptonstats}
\caption{The lepton sample statistics for the 1996-98 dataset.}
\vskip6pt
\begin{center}
\begin{tabular}{|l|c|c|c|c|}
\hline
 Channel & Sample & Efficiency & Purity & Dominant Background(s) \\ [-3pt]
         & Size   & $(|\cos\theta| < 0.8)$ &   & \\
\hline
\hline
$e^+e^-$ & 9419  &  87.3\%   &  98.6\%  & $\tau^+\tau^-$ (1.2\%)   \\
$\mu^+\mu^-$ &  7564  & 85.5\% & 99.8\% & $\tau^+\tau^-$ (0.2\%)   \\
$\tau^+\tau^-$ &  7088  & 78.1\% &  94.6\% & $\mu^+\mu^-$ (2.0\%), 
$2\gamma$ (1.7\%) \\
\hline
\end{tabular}
\end{center}
\end{table*}

\begin{figure}
\center
\psfig{figure=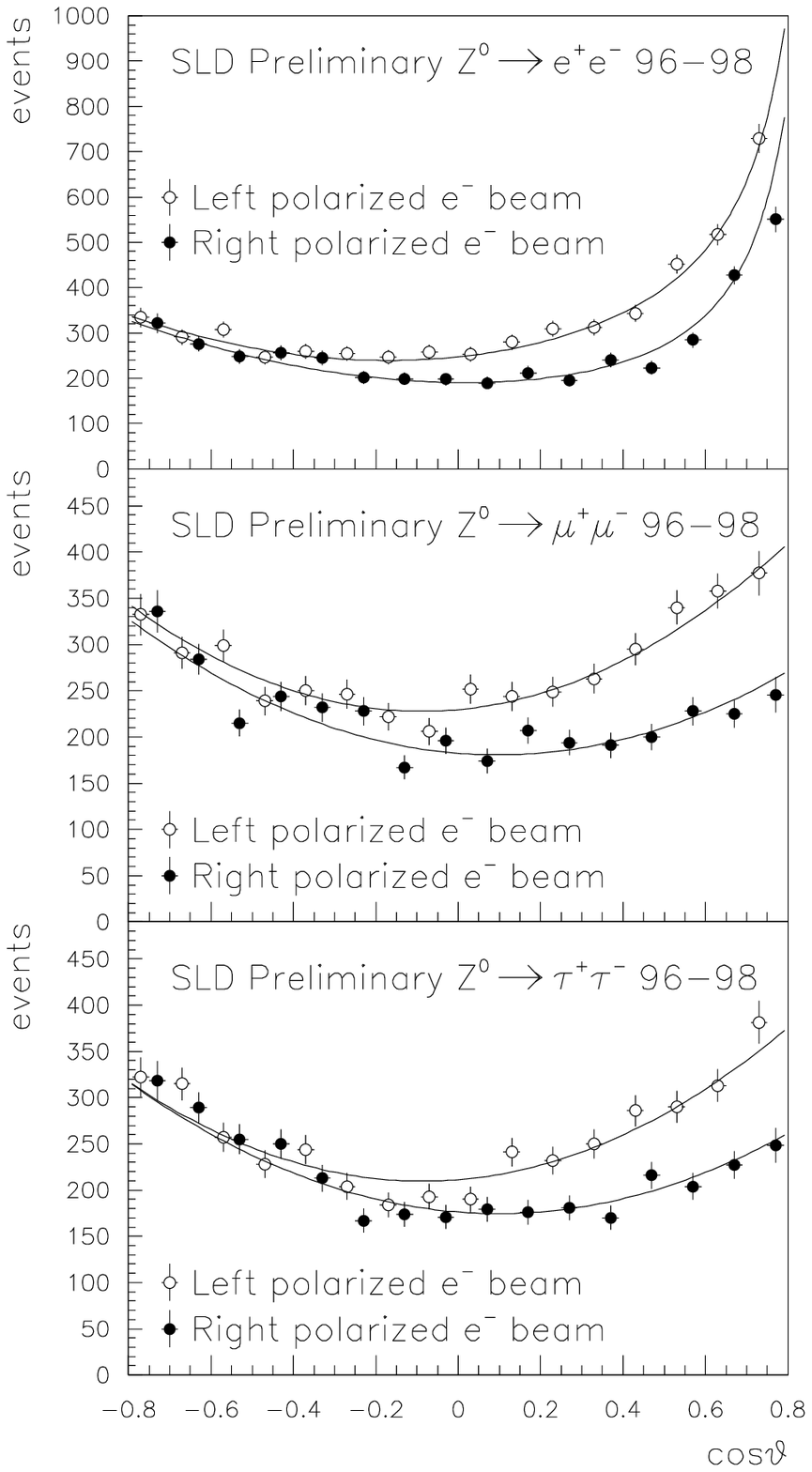,width=8cm}
\caption{Angular distributions for SLD lepton-pairs in the 1996-98
data sample, 
separately for left- and right-handed electron beams.}
\label{fig:leptons}
\end{figure}
                                                                                
An example of the resulting polar angle distributions for the 1996-98
dataset is shown in Fig. \ref{fig:leptons}.
The results of fits to the angular distributions, after
correcting for the small background contamination,
are summarized in Table \ref{table:leptonfits}. For the electron final state,
t-channel effects are incorporated into the angular
distribution, while for the tau final state, a
$\cos\theta$-dependent efficiency correction has been
applied to account for the correlation between visible
energy and net tau polarization.
Note that all channels provide information about $A_e$, which
comes in through the left-right cross section asymmetry.
The final-state lepton couplings are constrained by the
angular distributions. 

\begin{table*}
\caption{Results of the lepton asymmetry analysis for 1996-98 data}
\label{table:leptonfits}
\vskip6pt
\begin{center}
\begin{tabular}{|l|c|}
\hline
 Channel & $A_{lepton}$        \\
\hline
$\mu^+\mu^-$   & $A_{\mu}$=$0.128 \pm 0.022$ \\
$\tau^+\tau^-$ & $A_{\tau}$=$0.117 \pm 0.023$\\
$e^+e^-$,$\mu^+\mu^-$,$\tau^+\tau^-$ & 
$A_{e}$=$0.1497 \pm 0.0088 $ \\
\hline
\end{tabular}
\end{center}
\end{table*}

\begin{table*}
\caption{Combined results of the lepton asymmetry analysis for 1993-98 data}
\label{table:combolepton}
\vskip6pt
\begin{center}
\begin{tabular}{|l|c|c|}
\hline
 Channel & $A_{lepton}$ & $\sin^2\theta_W^{\sl eff}$ \\
\hline
$e^+e^-$       & $A_{e}$=$0.1504 \pm 0.0072$ & \\
$\mu^+\mu^-$   & $A_{\mu}$=$0.120 \pm 0.019$ & \\
$\tau^+\tau^-$ & $A_{\tau}$=$0.142 \pm 0.019$& \\
               & $A_{e \mu \tau}$=$0.1459 \pm 0.0063 $ & $0.2315 \pm 0.008$\\
\hline
\end{tabular}
\end{center}
\end{table*}
	
A summary of the (preliminary) combined results for the 1993-98
datasets is shown in Table \ref{table:combolepton}. The SLD result
for $A_\mu$
reflects the best single measurement, and is
competitive with the overall LEP
value $A_{\mu}^{LEP} = 0.145 \pm 0.013$~\cite{LEPEWG}.
A combination of the values in Table \ref{table:combolepton}, 
assuming lepton universality,
yields the value
\begin{eqnarray}
A_{lepton} & = & 0.1459 \pm 0.0063 \label{eqn:9398alepton}\\
\sin^2\theta_W^{\sl eff} & = & 0.2317 \pm 0.0008. 
\label{eqn:9398sin2twlepton}
\end{eqnarray}

\section{Summary of Results}

\begin{figure*}
\center
\vskip -1.1cm
  \epsfxsize=16.cm
  \epsfbox{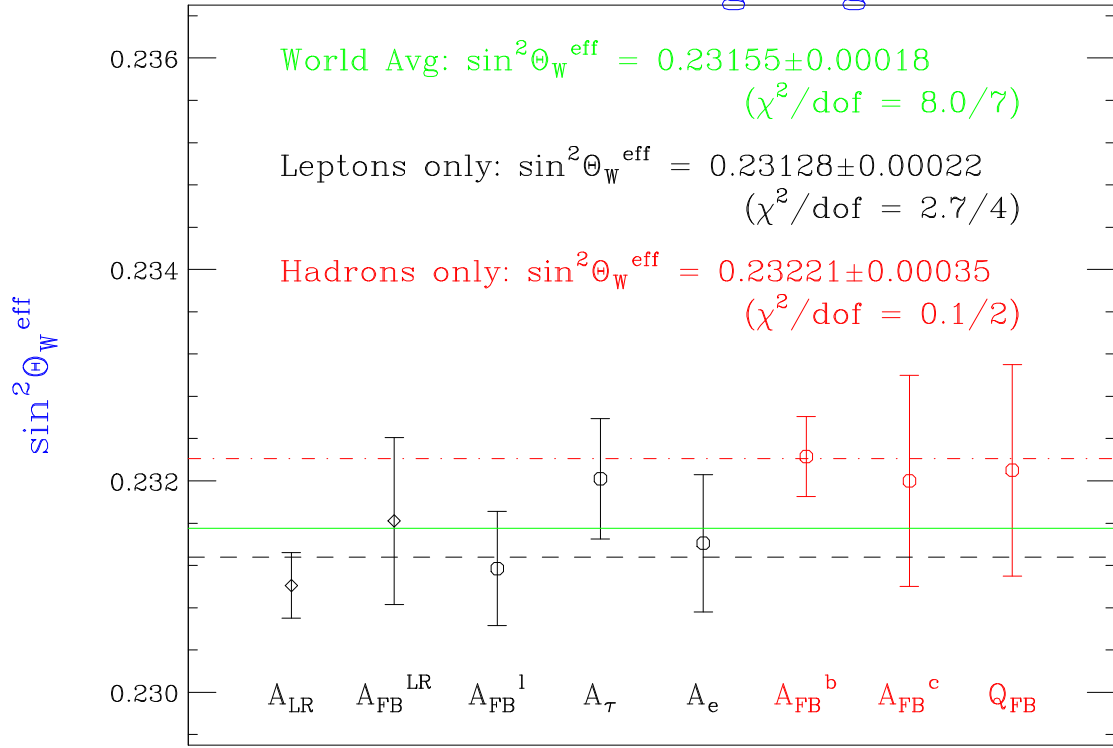}
  \caption{Status of the SLD and LEP $\sin^2\theta_W^{\sl eff}$
   measurements as of the XXIX$^{th}$ ICHEP conference.}
  \label{fig:sin2tw-all}
\end{figure*}

Combining these  results with previous results yields the preliminary 
SLD combined value of
\begin{eqnarray}
\sin^2\theta_W^{\sl eff} & = & 0.23110 \pm 0.00029.
\label{eqn:9398s2tw-combo}
\end{eqnarray}
Combining this with other recent results, we compute the SLD-LEP 
world average to be
\begin{eqnarray}
\sin^2\theta_W^{\sl eff} & = & 0.23155 \pm 0.00028.
\end{eqnarray}
These results are shown in Figure \ref{fig:sin2tw-all}. 
It is interesting to note that the world values for
$\sin^2\theta_W^{\sl eff}$ from lepton measurements (the results
presented here, the LEP leptonic averages $A_{FB}^l$ and $A_e$ , and
the LEP results of $A_\tau$ from $\tau$-polarization) are in excellent
agreement with each other, and differ by $\sim2.3\sigma$ from
those obtained from ``hadron-only'' measurements ($A_{FB}^b$,
$A_{FB}^c$, and $Q_{FB}$).


\section*{Acknowledgements}
I wish to thank Bruce Schumm, Peter Rowson, Morris Swartz, Toshinori
Abe, Tim Barklow, Mike Woods, and the rest of the SLD Electroweak working group
for their help in preparing my ICHEP presentation and this paper.

\end{document}